**Patient Specific Congestive Heart Failure Detection From Raw ECG signal**

Yakup KUTLU[1] , Apdullah YAYIK[2] , Esen YILDIRIM[1] , Mustafa YENİAD[1] ,Serdar YILDIRIM[1]

[1]Department of Computer Engineering İskenderun Technical University, Hatay, Turkey
[2]Turkish Army Forces, Turkey
[3]Department of Computer Engineering Yildirim Beyazit University, Ankara, Turkey

**Abstract**
In this study; in order to diagnose congestive heart failure (CHF) patients, non-linear second-order difference plot (SODP) obtained from raw 256 Hz sampled frequency and windowed record with different time of ECG records are used. All of the data rows are labelled with their belongings to classify much more realistically. SODPs are divided into different radius of quadrant regions and numbers of the points fall in the quadrants are computed in order to extract feature vectors. Fisher's linear discriminant, Naive Bayes, and artificial neural network are used as classifier. The results are considered in two step validation methods as general k-fold cross-validation and patient based cross-validation. As a result, it is shown that using neural network classifier with features obtained from SODP, the constructed system could distinguish normal and CHF patients with 100% accuracy rate.

**Keywords:**
Congestive heart failure, ECG, Second-Order Difference Plot, classification, patient based cross-validation.

**Introduction**

Heart failure is a stage when the heart unable to pump sufficient amount of blood that tissues need or just able to perform this with high filling pressures (Braunwald E, Zipes DP, Libby P 2004). Heart failure is a worldwide common disease affecting approximately 15 million people in the world. Its incidence increases with age. While there is 1-2% incidence between 50 and 60 year age group, it reaches 10% over the age of 75. Average 80% of all congestive heart failure (CHF) are seen in people over 65 years. 0.3% of men and 0.2% of women aged between 50 and 59 and 2.7% of man and 2.2% of women aged between 80 and 89 are exposed to CHF. The male / female ratio is calculated as 1 / 3. Each year in the United

States, causes of approximately 45000 patients death are declared as heart failure, and with each passing year, this number is increasing due to population aging and rising rates of cardiovascular disease-free survival. In addition, both medical expenses and loss of manpower negatively impacts on the economy (Topol & Califf 2007; Yayla 2010). Heart failure is a disease that could not be assessed easily in clinically. Early diagnosis and effective treatment of heart failure are obviously important for patients, and reduce the percentage of mortality and morbidity (Yayla 2010; Işler & Kuntalp 2007). Due to the fact that, appropriate way for treatment is possible if congestive heart failure disease is early diagnosed, automatic determination of CHF disease from ECG recordings is clinically very important. This study is an accomplished attempt to distinguish normal and CHF patients with a reliable classification system. Heart disease analysis is performed by Physical examination. However, definitive diagnosis of CHF could not be performed with Physical examination. So, in patients with suspected congestive heart failure, one or more diagnostic tests such as echocardiography, angiography, electrocardiography, chest X-ray film, brain (B-type) natriuretic peptide and stylistic trailblazer hormone, the N-terminal fragment, MR imaging are applied before making a decision (Yayla 2010; Işler & Kuntalp 2007; Kamath 2012a). In the literature, recently, by applying different signal processing and analysis methods, it is tried to advance effective computer-aided diagnosis methods (Işler & Kuntalp 2007; Kamath 2012b; Kannathal et al. 2006; Thuraisingham 2010; Karmakar et al. 2009; D. 2009; Engin 2007; Maurice E.Cohen 1996; Dabanloo et al. 2010; Işler Y n.d.). Some of them considered Heart Rate Variability (HRV) of ECG signals (Işler & Kuntalp 2007; Kamath 2012b; Işler Y n.d.; Kannathal et al. 2006; Thuraisingham 2010). In both cases, hidden important information (such as wavelet, eigenvector methods, Poincare, RR interval, etc.) in the ECG record are revealed and used as feature vectors for detecting CHF patients. Işler and Kuntalp (Işler & Kuntalp 2007) used HRV to analyse CHF disease and they used many techniques to extract features such as time frequency domain features, frequency domain features, point care plot features, and patient information. Kamath (Kamath 2012b) reported the central tendency measure of the R-R interval and showed the efficacy of redial distance of the Teager energy scatter plot to distinguish CHF from normal subjects. A measure of complex correlation to quantify temporal variability in the Poincare plot was introduced by Karmakar et al. (Karmakar et al. 2009). Thuraisingham (Thuraisingham 2010) used RR interval to analyse SODP of RR intervals. And he introduced a classification system which employs a statistical procedure. Cohen et al. (Maurice E.Cohen 1996) analysed SODPs and Central tendency measure using CHF and HRV data. Some of them considered beat based methods (Kannathal et al. 2006; Karmakar et al. 2009; D. 2009; Engin 2007). Furthermore, different methods (such as expert system, KNN, fuzzy, neuro-fuzzy, etc.) have been used as classifier. Poincare plot and the SODP are commonly used methods as non-linear analysis of the components in biomedical signals (Işler Y n.d.; Thuraisingham 2010; Karmakar et al. 2009; Maurice E.Cohen 1996; Dabanloo et al. 2010). It is also known that poincare plot is an example of chaotic system (Maurice E.Cohen 1996). While poincare map reveals the relations of the consecutive point with each other, SODP reveals the relationship of consecutive difference values with each other (Kamath 2012b; Thuraisingham 2010; Maurice E.Cohen 1996). When researcher used RR interval series to obtain features, they constructed the system by using R-R interval obtained holter record. It was long-term record. Therefore many researchers still try to increase their success and reliability by developing new approaches. They try to find less computational complexity and much useful system needed less data information. In this study, the goal is to construct a system that could be worked in real time and fast and high performance. So, whole new meaningful features are tried to extract from SODP measurement using raw ECG record in order to distinguish CHF and normal patients. And it is focused two objectives that to show reliability of sampling frequency and window size of pattern in analysis. In present study, the ECG pattern with CHF and normal ECG records are achieved from the Physiobank database. Neural Network, Naive Bayes, and Linear Discriminant algorithm are used to construct classification system. Then the results are considered in two step validation methods as general k-fold cross-validation and patient based cross-validation.

In the following section, the data acquisition, pre-processing steps, method of feature extraction, the classifiers and performance measures of classifier are presented. In Section 3, the results and discussion are given. The conclusion of the study is presented in Section 4.

**Material and methods**

This section introduced the data acquisition, process of data preparation and feature extraction methods used in the proposed recognition system. Figure 1 shows the general block diagram of the constructed system.

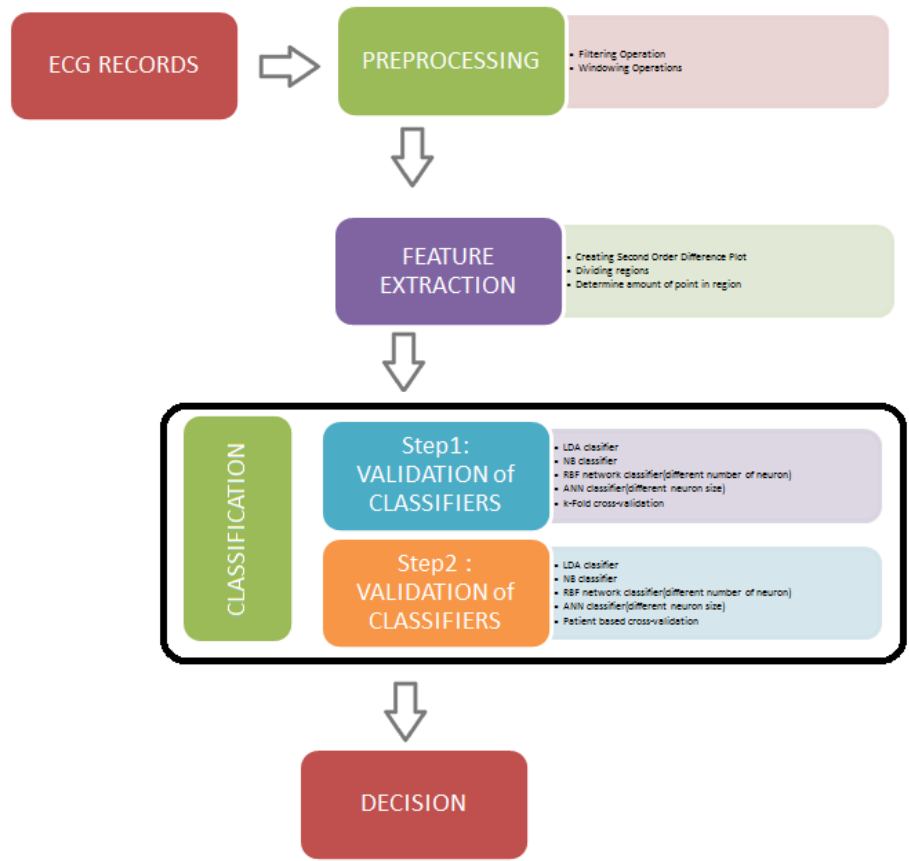

Figure 1. Block diagram of the proposed system.

*Data acquisition*

ECG data are obtained freely from PhysioNet web site as free of charge (A.L. Goldberger and coworkers 2000). In this study, CHF Database and Normal Sinus Rhythm Database are used. CHF database are 24 hour ECG records, obtained from 256 Hz sampling rate and received from 15 patients changing between 22 and 71 years old. Normal sinus rhythm database are 24 hour ECG records, obtained from 256 Hz sampling rate and received from 18 non-patient volunteers changing between 20 and 50 years old. ECG records have Power line interference and Baseline wander effects because of respiration. Baseline wander and Power line interference consist of low frequency components and high frequency components, respectively. The records are filtered with two median filters to eliminate the baseline wander and with a notch filter to eliminate power-line frequency (Suri et al. 2007). In all other processing, the filtered signals are used. In order to examine amount of information in a window time, the records are divided into four different window time that are 10 sec, 7 sec, 5 sec and 3 sec. for 10 sec of window time, 150 windowed sample are extracted from record. Total of 4950 windowed samples are extracted for analyse. Similarly, for other window time, windowed samples are extracted from record. The samples of windowed records are shown in Figure 2. The normalization process is applied by dividing each beat to absolute maximum value (Duda & Hart n.d.; Bishop & others 1995) for each sample, before extracting features from raw ECG data records.

*Second-Order Difference Plot (SODP)*

SODP is a feature extraction method that obtains form time domain information. Implementation of SODP is very easy. The method of SODP is used both to provide independent feature extraction tools and to be used as complementary method to verify the frequency domain results (Thuraisingham 2010; Maurice E.Cohen 1996) . SODP has meaningful information about ECG (Kamath 2012b; Thuraisingham 2010). If X(t) is the ECG signal, SODP is formed by [X (t+1)-X (t)], and [X (t+2)-X (t+1)] points on the plot. In other words, SODP includes scattering of consecutive difference values of points in ECG signal. Thus, the statistical situation of consecutive differences can be observed. Figure 3 shows the Second-order difference plot of CHF and Normal ECG signals.

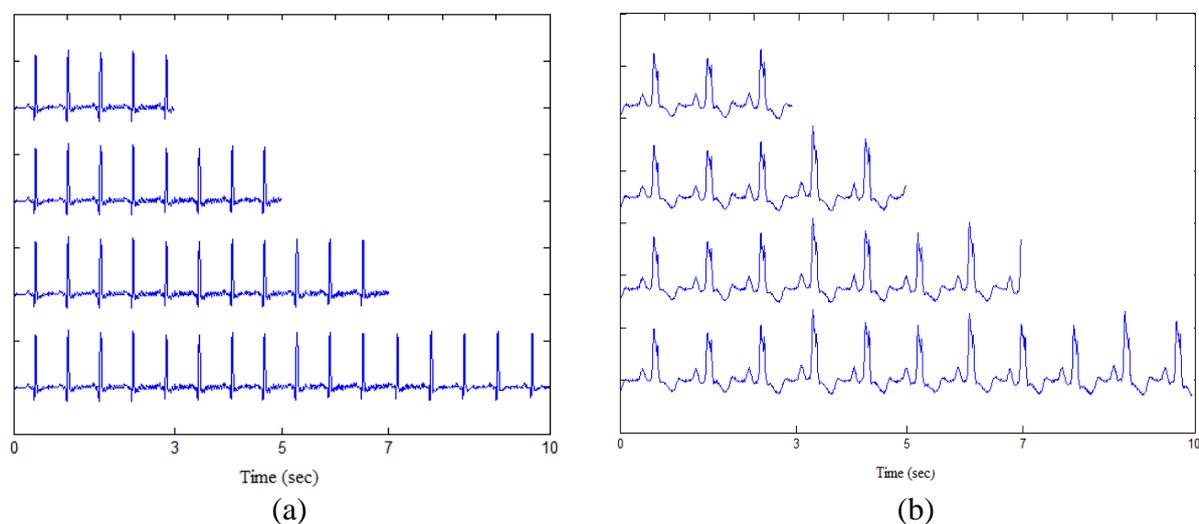

(a)                           (b)

Figure 2. The windowed records (respectively, 10 sec, 7 sec, 5 sec and 3 sec) a)Normal ECG record b) CHF record

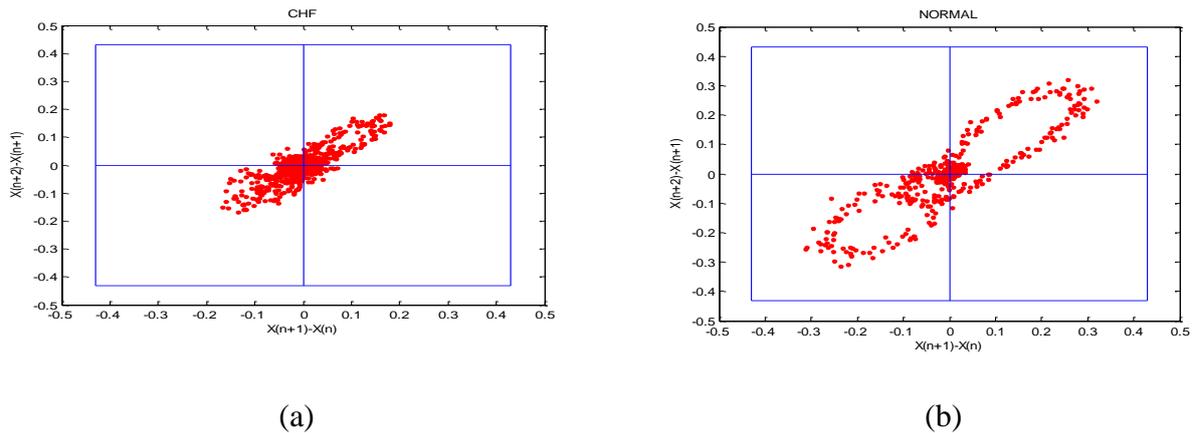

(a) (b)

Figure 3. A sample of the Second-order difference plot for (a) CHF and (b) Normal ECG signal

*Feature Extraction*

Feature is information of the pattern. Therefore, the information is tried to extract from second-order difference plot of the CHF and Normal ECG record. The SODP is used to extract information about signals. The SODP is a figure of two-dimensional Cartesian system. The axes of a two-dimensional Cartesian system divide the quadrants which are four infinite regions numbered from 1st to 4th, each bounded by two half-axes. The region number and two coordinates sings are I ( + , + ), II ( − , + ), III ( − , − ) and IV ( + , − ). In the studies, SODPs are divided into different radius of circle regions in order to extract features (Thuraisingham 2010; Kamath 2012b). Second-order difference plot's four regions of a quadrant are shown below (Figure 4). It shows the region divided by circles centered at the origin. There are four quadrants of a Cartesian coordinate system. Each quadrant has four region divided circle. Each region shows the increasing number of points in SODP. There are sixteen different regions (four regions of four quadrants). The numbers of points are calculated for each region and used as a feature vectors. All quadrant regions contain valuable information to classify SODP points. In the second region ( − , + ) and the fourth region ( + , − ) represent balanced increasing and decreasing, the first region ( + , + ) and the third Region ( − , − ) represent the continuous increasing and decreasing (Kamath 2012b).

*Performance Measurements*

In this study, total of 33 ECG records (15 CHF and 18 normal ECG) is used. The records are windowed as 3 sec, 5 sec, 7 sec, 10 sec. 150 windowed record are used for each record. 16 features are extracted from each window. Total of 4950 feature vectors are extracted for each window. Performances are evaluated by using LDA, NB, network, ANN classifiers. The performances are validated by two methods: general k-fold cross-validation and patient based cross-validation.

*General k-fold Cross-validation*

Cross-validation is also known rotation estimation. It is a method to determine how the results of a statistical analysis will generalize to a new data set. In this method, all data set is randomly separated k equal subsets. One subset is used as validation and all other subsets are used as train. This step iterated for k times leaving one different fold for evaluation each time. This validation method is performed for better approximating error (Duda & Hart n.d.).

*Patient Based Cross-validation*

Patient based cross-validation is a method that is similar to k-fold cross-validation. Prepared data sets are obtained from M different people. Each data vector are labelled both as information of disease and patient numbers. In the patient based cross-validation, feature vectors that belong to only one person are used as validation and all other data are used as training. The step is iterated M times leaving one different person's data for evaluation each time. M different test set from different people are classified. Correctly classification rate of individual data is used to determine the performance of the system. According to rate of having CHF the system decides whether or not the patient has any disease. When the rate is over 0.50 the decision is positive. When the rate is less than 0.50 the decision is negative. According to being CHF and Normal record, the results are determined. The performance of the system is assessed as rate of all correctly classified patients.

**Result and Discussion**

This study is carried out with using Core2Due 2.4 GHz processor, PC that has 2 GB of memory and MATLAB package program. Total of 33 ECG records (15 CHF and 18 normal ECG) are used. In literature both frequencies are used to analyse. So both normal and CHF patient's ECG records with 256 Hz sampling frequency are considered in this study for analysing. The CHF is a pattern type heart disease so the analysed data should consist of at least three peaks. Therefore, in order to examine amount of information in a window of raw ECG pattern, the records are divided into four different window time (10 sec, 7 sec, 5 sec and 3 sec). 4950 windowed samples of raw ECG records are arranged for each window time. These data sets are examined separately. SODP is used to extract information from ECG signal in this study. SODP is divided into four circles with different radius for CHF and normal data as shown in Figure 7. There are four region of a quadrant. The number of points that fall in the region of quadrants is calculated. This information is used as feature vectors. 16-dimensional feature vectors are generated and the classifier performances are analysed.

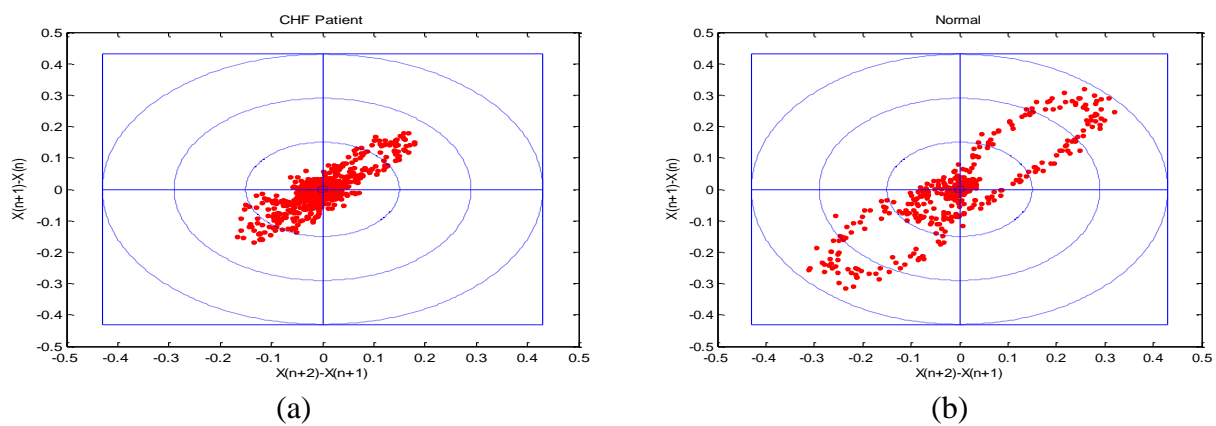

Figure 7. SODP that is divided into four radius circle region (a) CHF patient's ECG signal (b) Normal ECG signal

The validation step consists of two parts that are performed as general 10-fold cross-validation and patient based cross-validation. At first, the feature set is classified without information of candidate. So all windowed data are only labelled as CHF and Normal and used to construct a classification system with high accuracy. LDA, NB, MLP, are used as classifier. The records with two different sampling frequencies and four different windows for each record and four different classifier models are used in the analysis. Different case (combination of two sampling rate, four window time, classifier with different parameter such as neuron size of MLP) are considered.

| Classifier | Window Time | SEN | SEL | SPE | ACC |
|---|---|---|---|---|---|
| LDA | 3sec | 95.72 | 96.23 | 95.08 | 96.10 |
|  | 5sec | 96.10 | 97.06 | 96.64 | 96.55 |
|  | 7sec | 97.72 | 97.80 | 97.43 | 97.81 |
|  | 10sec | 98.04 | 98.32 | 97.79 | 98.15 |
| NB | 3sec | 90.36 | 93.85 | 92.45 | 92.26 |
|  | 5sec | 90.40 | 95.00 | 93.78 | 92.91 |
|  | 7sec | 90.49 | 95.41 | 94.26 | 93.17 |
|  | 10sec | 90.22 | 95.89 | 94.82 | 93.31 |

Table 1. The performance measure of LDA and NB classifiers.

Sensitivity, selectivity, specificity, overall accuracy and correct classification rate are used as performance measure of classifiers. The general 10-fold cross-validation method is used in this step. The 10 equal subsets are prepared from data set. One subset is used as validation and all other subsets are used as train set. The process is iterated for 10 times. The performance measurements (sensitivity, selectivity, specificity and accuracy) for different window time, and for two sampling frequency are shown in Table 1 and 2 for classifiers of LDA, NB and MLP, respectively. The best performance could be taken by using window time of 10 sec. The best overall accuracy of classifiers are 98.15%, 93.31% and 99.96%, respectively LDA, NB MLP classifiers. The system was planned to build using a classifier with the best performance. For this purpose, the classifiers with the best results that obtained with the above Tables, the second step of validation accomplished. The second step of validation was patient based cross-validation. The goal of this step, the candidate subject was separated CHF subjects from healthy subjects with high performance. In this step, the feature vectors are obtained from 33 (18 normal and 15 CHF patient) different candidates. Each data vectors are labelled both as information of disease and patient numbers. In the patient based cross-validation, 150 feature vectors that belongs to only one person are used as test set and all other data are used as training set. This process is iterated 33 times for all patients' data. 33 different test set from different people are classified as CHF or Normal. According to rate of having CHF which is correctly classification rate of individual data, the system decides whether or not the patient has CHF. When the rate of having CHF is over 0.50 the decision is assessed as positive. When the rate of having CHF is less than 0.50 the decision is assessed as negative. According to being CHF and Normal record, the results are determined. The diagnosis of the patient is determined by looking to the value of rate. If rate is over 0.50, the person is a patient with CHF. Otherwise, the person is a normal patient. Table 3 shows the result of a classifier. The result indicates that whether the decision is correct or not. The error measurement of the system is assessed as rate of all misclassified CHF and Normal patients. The misclassification performance measurements of systems are shown 2.94, 0.00, 2.94 for LDA, NB, MLP classifiers. It can be released seen that MLP classifier has 0% of misclassification rate which is the pest performance. The comparison of the proposed systems with similar systems is difficult because of the varieties in the classification techniques, feature extraction techniques, data used in system, and performance measurements.

| Window Time | Neuron Size | SEN | SEL | SPE | ACC |
|---|---|---|---|---|---|
| 3sec | 3 | 99.87 | 99.74 | 99.66 | 99.16 |
|  | 5 | 99.82 | 99.70 | 99.65 | 99.76 |
|  | 7 | 99.91 | 99.85 | 99.82 | 99.88 |
|  | 9 | 99.64 | 99.78 | 99.73 | 99.72 |
|  | **Average** | **97.81** | **99.77** | **99.72** | **98.88** |
| 5sec | 3 | 99.78 | 99.81 | 99.78 | 99.80 |
|  | 5 | 99.64 | 99.85 | 99.82 | 99.76 |
|  | 7 | 99.60 | 99.81 | 99.78 | 99.72 |
|  | 9 | 99.87 | 99.81 | 99.78 | 99.84 |
|  | **Average** | **99.72** | **99.82** | **99.79** | **99.78** |
| 7sec | 3 | 99.82 | 99.93 | 99.91 | 99.88 |
|  | 5 | 99.78 | 99.93 | 99.91 | 99.86 |
|  | 7 | 99.78 | 100 | 100 | 99.35 |
|  | 9 | 98.93 | 100 | 100 | 99.52 |
|  | **Average** | **99.58** | **99.96** | **99.96** | **99.65** |
| 10sec | 3 | 99.64 | 99.89 | 99.87 | 99.78 |
|  | 5 | 99.87 | 99.93 | 99.91 | 99.90 |
|  | 7 | 99.87 | 100 | 100 | 99.94 |
|  | 9 | 100 | 99.93 | 99.91 | 99.96 |
|  | **Average** | **99.84** | **99.94** | **99.92** | **99.89** |

Table 2. The performance measure of MLP classifier.

Nonetheless, some conclusions can be drawn. Işler and Kuntalp (Işler & Kuntalp 2007) used R-R interval series and used many techniques to extract features such as time frequency domain features, frequency domain features, point care plot features, and patient information. Genetic algorithm is used to find best subsets of these feature vectors. They reported the maximum performances with sensitivity of 100% and specificity of: 94.74%. But the methods have high calculation complexity. Kamath (Kamath 2012b) reported almost 100% classification rate of CHF subject using k-nearest neighbor classifier analyzing R-R interval series using the central tendency measure of the plots. Thuraisingham (Thuraisingham 2010) reported Central tendency measure to extract features using SODP of RR intervals. He reported as almost 100% of accuracy using a recognition system which utilizes a statistical method to distinguish CHF from normal patients. When the researcher used RR interval series to extract features, they need all R-R interval of holter record to analyse. It is long-term record and is required high computational time. The proposed system could separate CHF form Normal candidates with excellent performance. Furthermore, prepared system used only few part of record to detect CHF patient instead of the whole ECG record.

|  | Patient ID | Total of Sample | Result of Classifier CHF | Result of Classifier Normal | Rate of being CHF | Decision of System | Decision |
|---|---|---|---|---|---|---|---|
| Records with CHF | 1 | 150 | 150 | 0 | 1.00 | Positive | T |
| | 2 | 150 | 39 | 111 | 0.26 | **Negative** | **F** |
| | 3 | 150 | 133 | 17 | 0.89 | Positive | T |
| | 4 | 150 | 150 | 0 | 1.00 | Positive | T |
| | 5 | 150 | 72 | 78 | 0.48 | **Negative** | **F** |
| | 6 | 150 | 150 | 0 | 1.00 | Positive | T |
| | 7 | 150 | 149 | 1 | 0.99 | Positive | T |
| | 8 | 150 | 49 | 101 | 0.33 | **Negative** | **F** |
| | 9 | 150 | 150 | 0 | 1.00 | Positive | T |
| | 10 | 150 | 150 | 0 | 1.00 | Positive | T |
| | 11 | 150 | 150 | 0 | 1.00 | Positive | T |
| | 12 | 150 | 147 | 3 | 0.98 | Positive | T |
| | 13 | 150 | 86 | 64 | 0.57 | Positive | T |
| | 14 | 150 | 149 | 1 | 0.99 | Positive | T |
| | 15 | 150 | 150 | 0 | 1.00 | Positive | T |
| Normal Records | 16265 | 150 | 3 | 147 | 0.02 | Negative | T |
| | 16272 | 150 | 3 | 147 | 0.02 | Negative | T |
| | 16273 | 150 | 0 | 150 | 0.00 | Negative | T |
| | 16420 | 150 | 2 | 148 | 0.01 | Negative | T |
| | 16483 | 150 | 0 | 150 | 0.00 | Negative | T |
| | 16539 | 150 | 0 | 150 | 0.00 | Negative | T |
| | 16773 | 150 | 0 | 150 | 0.00 | Negative | T |
| | 16786 | 150 | 0 | 150 | 0.00 | Negative | T |
| | 16795 | 150 | 146 | 4 | 0.97 | **Positive** | **F** |
| | 17052 | 150 | 6 | 144 | 0.04 | Negative | T |
| | 17453 | 150 | 1 | 149 | 0.01 | Negative | T |
| | 18177 | 150 | 0 | 150 | 0.00 | Negative | T |
| | 18184 | 150 | 0 | 150 | 0.00 | Negative | T |
| | 19088 | 150 | 0 | 150 | 0.00 | Negative | T |
| | 19090 | 150 | 20 | 130 | 0.13 | Negative | T |
| | 19093 | 150 | 0 | 150 | 0.00 | Negative | T |
| | 19140 | 150 | 1 | 149 | 0.01 | Negative | T |
| | 19830 | 150 | 0 | 150 | 0.00 | Negative | T |

Table 3 The best performance measure of patient based classification (for window time of 10 sec, for 250 Hz sampled frequency and for LDA classifier)

So that computational time is low enough. And in proposed study, using 20 minute of ECG recording the system has 100% correct identification rate of CHF patient and it added a new approach in this area. In each study in the literature, however, it should be noticed that different data types have been used, so, it is actually difficult to compare the results in an objective and fair way. The results demonstrate that the proposed method is more successful in discriminating the records of CHF and normal patient.

**References**

A.L. Goldberger and coworkers, 2000. PhysioBank, PhysioToolkit, and PhysioNet: Components of a New Research Resource for Complex Physiologic Signals. *Circulation 101(23):e215-e220*. Available at: http://circ.ahajournals.org/cgi/content/full/101/23/e215.

Bishop, C.M. & others, 1995. Neural networks for pattern recognition.

Braunwald E, Zipes DP, Libby P, B.R., 2004. *Braunwald's Heart Disease A Textbook of Cardiovascular Medicine* 7th Editio., Philadelphia, USA: Saunders.

D., U.E., 2009. Adaptive neuro-fuzzy inference system for classification of ECG signals using Lyapunov exponents. *Computer Methods and Programs in Biomedicine.*, pp.313–321.

Dabanloo, N.J. et al., 2010. Application of Novel Mapping for Heart Rate Phase Space and Its Role in Cardiac Arrhythmia Diagnosis. *Computing in Cardiology*, pp.209–212.

Duda, R.O. & Hart, P.E., Pattern Classification.pdf.

Engin, M., 2007. Spectral and wavelet based assessment of congestive heart failure patients. *Computers in Biology and Medicine*.

Işler Y, K.M., Konjestif Kalp Yetmezligi Teshisi için Kalp Hızı Degiskenligi Analizinde Dalgacık Entropisinin Etkisi. In *IEEE 14th Signal Processing and Communications Applications Conference*. pp. 1–4.

Işler, Y. & Kuntalp, M., 2007. Combining classical HRV indices with wavelet entropy measures improves to performance in diagnosing congestive heart failure. *Computers in biology and medicine*, 37(10), pp.1502–10. Available at: http://www.ncbi.nlm.nih.gov/pubmed/17359959 [Accessed June 22, 2013].

Kamath, C., 2012a. A new approach to detect congestive heart failure using sequential spectrum of electrocardiogram signals. *Medical engineering & physics*, 34(10), pp.1503–9. Available at: http://www.ncbi.nlm.nih.gov/pubmed/22459502 [Accessed December 8, 2013].

Kamath, C., 2012b. A new approach to detect congestive heart failure using Teager energy nonlinear scatter plot of R-R interval series. *Medical engineering & physics*, 34(7), pp.841–8. Available at: http://www.ncbi.nlm.nih.gov/pubmed/22032833 [Accessed June 22, 2013].

Kannathal, N. et al., 2006. Cardiac state diagnosis using adaptive neuro-fuzzy technique. *Medical Engineering & Physics*, 28(8), pp.809–815.

Karmakar, C.K. et al., 2009. Novel feature for quantifying temporal variability of Poincare Plot: a case study. In *Computers in Cardiology, 2009*. pp. 53–56.

Maurice E.Cohen, D.L.H. and P.C.D., 1996. Applying Continuous chaotic Modeling to Cardiac Signal Analysis. *Engineering in Medicine and Biology*, (2), pp.97–102.

Suri, J. et al., 2007. *Advances in cardiac signal processing*, Springer.

Thuraisingham, R.A., 2010. A Classification System to Detect Congestive Heart Failure Using Second-Order Difference Plot of RR Intervals. *Cardiology research and practice*, 2009, p.807379. Available at: http://www.pubmedcentral.nih.gov/articlerender.fcgi?artid=2842886&tool=pmcentrez&rendertype=abstract [Accessed June 22, 2013].

Topol, E.J. & Califf, R.M., 2007. *Textbook of cardiovascular medicine*, Lippincott Williams & Wilkins.

Yayla, B., 2010. *Kalp yetmezlikli hastalarda adrenomedullin ve probnp düzeylerinin ekokardiyografi ile ilişkisi*. İstanbul Göztepe Araştırma ve Geliştirme hastanesi.